\documentclass[a4paper]{PoS}
\pdfoutput=1

\usepackage{amsmath}
\usepackage{amssymb}
\usepackage{amsthm}
\usepackage{multirow}

\def\tanb{\ensuremath{\tan\beta}}

\def\half{\ensuremath{\frac{1}{2}}}
\def\beq{\begin{equation}}
\def\eeq{\end{equation}}
\def\bea{\begin{eqnarray}}
\def\eea{\end{eqnarray}}

\newcommand{\hobs}{H_{\rm obs}}

\newcommand{\hpm}{H^\pm}
\newcommand{\hmp}{H^\mp}

\newcommand{\wpm}{W^\pm}
\newcommand{\wmp}{W^\mp}

\title{Signatures of the Type-I 2HDM at the LHC}

\ShortTitle{Signatures of the Type-I 2HDM at the LHC}

\author{Rikard Enberg,$^a$ William Klemm,$^{ab}$ Stefano Moretti$^c$ and \speaker{Shoaib Munir}$^d$\\
        \llap{$^a$}Department of Physics and Astronomy, Uppsala University, \\
Box 516, SE-751 20 Uppsala, Sweden\\
        \llap{$^b$}School of Physics \& Astronomy, University of Manchester, \\
Manchester M13 9PL, UK \\
       \llap{$^c$}School of Physics \& Astronomy, University of Southampton, \\
       Southampton SO17 1BJ, UK \\
       \llap{$^d$}School of Physics, Korea Institute for Advanced Study, \\
Seoul 130-722, Republic of Korea \\
        E-mail: \email{rikard.enberg@physics.uu.se}, \email{william.klemm@physics.uu.se}, \email{s.moretti@soton.ac.uk}, \email{smunir@kias.re.kr}}

\abstract{One of the simplest extensions of the Standard Model (SM) is the two-Higgs-doublet model (2HDM), which contains two neutral Higgs bosons, in addition to a 125\,GeV one, and a charged pair. At the Large Hadron Collider (LHC), gluon-induced processes are generally the most important modes for the resonant production of the SM-like Higgs boson as well as its pair-production, and it is generally considered to be the case also for an additional neutral Higgs boson possibly existing in nature. We show that for certain parameter configurations in the Type-I 2HDM, electroweak pair-production of the neutral Higgs states can dominate over the QCD-initiated production. Moreover, it is possible for the pair-production of the charged Higgs state along with a neutral one, which can only take place electroweakly, to have a substantial cross section. We delineate such 2HDM parameter space regions through its comprehensive numerical scanning, requiring their consistency with the most relevant theoretical and experimental constraints. We also highlight some specific di-Higgs signatures that can be probed at the LHC in order to establish the Type-I 2HDM as the underlying new physics model.}

\FullConference{Corfu Summer Institute 2018 "School and Workshops on Elementary Particle Physics and Gravity"\\
		(CORFU2018)\\
		31 August - 28 September, 2018\\
		Corfu, Greece}

\begin{document}

\section{The Type-I 2HDM}

The 2HDM is obtained by augmenting the complex scalar doublet, $\Phi_1$, of the SM by
 another doublet, $\Phi_2$, which alters the dynamics of electroweak (EW) symmetry-breaking. 
Three out of the eight degrees of freedom in the Higgs sector of the model lend masses to the EW gauge bosons, and the remaining five manifest themselves as physical states. These states include two scalars ($h$ and $H$, with $m_h<m_H$),
 a pseudoscalar ($A$), and a $\hpm$ pair. The most general CP-conserving scalar potential of the 2HDM can be written as
\begin{equation}
\begin{split}
\mathcal{V}_{\rm 2HDM}  &= m_{11}^2\Phi_1^\dagger\Phi_1+ m_{22}^2\Phi_2^\dagger\Phi_2
-[m_{12}^2\Phi_1^\dagger\Phi_2+ \, \text{h.c.} ] \\
& +\half\lambda_1(\Phi_1^\dagger\Phi_1)^2
+\half\lambda_2(\Phi_2^\dagger\Phi_2)^2
+\lambda_3(\Phi_1^\dagger\Phi_1)(\Phi_2^\dagger\Phi_2)
+\lambda_4(\Phi_1^\dagger\Phi_2)(\Phi_2^\dagger\Phi_1) \\
& +\left\{\half\lambda_5(\Phi_1^\dagger\Phi_2)^2
+\, \text{h.c.}\right\}\,.
\label{eq:2hdmpot}
\end{split}
\end{equation}
When the EW symmetry is broken, the fields $\Phi_1$ and $\Phi_2$ in the above potential are expanded around their vacuum expectation values $v_1$ and $v_2$, respectively, as
\begin{equation}
\Phi_1=\frac{1}{\sqrt{2}}\left(\begin{array}{c}
\displaystyle \sqrt{2}\left(G^+\cos\beta -H^+\sin\beta\right)  \\
\displaystyle v_1-h\sin\alpha+H\cos\alpha+\mathrm{i}\left( G\cos\beta-A\sin\beta \right)
\end{array}
\right),
\end{equation}
\begin{equation}
\Phi_2=\frac{1}{\sqrt{2}}\left(\begin{array}{c}
\displaystyle \sqrt{2}\left(G^+\sin\beta +H^+\cos\beta\right)  \\
\displaystyle v_2+h\cos\alpha+H\sin\alpha+\mathrm{i}\left( G\sin\beta+A\cos\beta \right)
\end{array}
\right),
\end{equation}
with $G$ and $G^\pm$ being the Goldstone bosons, $\alpha$ being the mixing angle of the CP-even interaction states, and $\tanb\equiv v_1/v_2$. Using the minimisation conditions, the mass parameters $m_{11,22}^2$ appearing in the Higgs potential can be replaced by $v_{1,2}$, while the quartic couplings $\lambda_{1-5}$ can be traded for the parameter $\sin(\beta-\alpha)$ and the masses of the four Higgs bosons. This results in a total of seven free parameters in the 2HDM: $m_h,\,m_H,\,m_A,\,m_{H^\pm},\,m_{12}^2,\,\tanb$ and $\sin(\beta-\alpha)$.

In principle, the Yukawa couplings of the fermions are also free parameters of the model. However, if both $\Phi_1$ and $\Phi_2$ couple to all the fermions, they can mediate flavour-changing neutral currents (FCNCs) at the tree level. The simplest way to avoid dangerously large FCNCs is to enforce a $Z_2$ symmetry on the model Lagrangian, which implies that only one of the doublets couples to a given type of fermions \cite{Glashow:1976nt,Paschos:1976ay}. The $m_{12}^2$ term in the Higgs potential softly breaks this symmetry. The 2HDM can be classified into several {\it Types} depending on the charge assignment of the fields under the $Z_2$ symmetry. The Type-I 2HDM is obtained if all the quarks and charged leptons couple only to $\Phi_2$, by imposing $\Phi_1 \to -\Phi_1$.  

\section{EW production of di-Higgs states at the LHC}

In the context of the LHC, most studies in the literature have conventionally focussed on the QCD-induced production of Higgs bosons in the 2HDM, whether single or multiple (see, e.g., Ref. \cite{Arhrib:2009hc,Hespel:2014sla} for a review). For the production of Higgs bosons pairs in gluon-initiated processes, there are two modes of relevance: $s$-channel processes involving a neutral Higgs (or an off-shell $Z$) boson in the propagator, and box diagrams involving heavy fermion loops. The former, known as gluon-fusion process, is by far the dominant mode for the resonant production of an {\it SM-like} Higgs boson at the LHC. In models beyond the SM also, it can dominate strongly in the production of pairs of Higgs bosons through an intermediate heavier Higgs state, if the triple-Higgs couplings involved are sufficiently large. Furthermore, the Yukawa couplings of the $b$-quarks can be quite sizeable for certain parameter space configurations in these models, making $b\bar{b}$-fusion another crucial production mode for single (intermediate) Higgs bosons. In essence though, the fact that each incoming $b$ is a sea-quark that results from a (double) gluon splitting, makes this channel also intrinsically $gg$-induced. 

In some recent studies \cite{Enberg:2016ygw,Arhrib:2017wmo,Enberg:2018pye}, we have analysed the prospects of di-Higgs production instead in $q\bar q^{(')}$-induced processes, where $q$ represents predominantly the valence $u$- and $d$-quarks, at the $\sqrt{s} = 13$\,TeV LHC. These studies aimed at scrutinising whether the cross sections for this EW production of (some of) the di-Higgs states can exceed those from the QCD-initiated processes. As for the charged di-Higgs states (i.e, states comprising of the $\hpm$ and a neutral Higgs boson - the production of which is precluded to the $gg$-induced processes), we looked to establish if their EW production can be strong enough to make them potentially accessible at the current or future LHC Runs. Importantly, the accessibility of various di-Higgs states is essential for probing the corresponding triple-Higgs couplings appearing in the Lagrangian of the Type-I 2HDM. In fact, as their production is often mediated by the $W$ and $Z$ bosons, di-Higgs states can provide sensitivity even to the Higgs-Higgs-gauge couplings.

\section{Analysis methodology}

For each of the analyses that will be described in the following sections, we first performed numerical scanning of the Type-I 2HDM parameter space, using the 2HDM Calculator (2HDMC)~\cite{Eriksson:2009ws}. In the 2HDM, either one of $h$ and $H$ can play the role of the {\it SM-like} Higgs boson, $\hobs$, observed at the LHC \cite{Aad:2012tfa,Chatrchyan:2012xdj}. We therefore analysed two separate cases, with the mass of $h$ in one, and that of $H$ in the other, fixed to 125\,GeV. As for the other free parameters, the following scanned ranges were uniform across all the studies
\begin{center}
		$\sin(\beta-\alpha)=-1$ -- 1\,;~~$m_{12}^2= 0$ -- $m_A^2\sin\beta\cos\beta$\,;~~$\tanb= 2$ -- 25\,, \\
\end{center}
while the ranges of the masses of the remaining Higgs bosons were chosen depending on the focus of a given study. During the scanning process, each sampled point was required to satisfy the basic theoretical conditions of unitarity, perturbativity, and stability of the Higgs potential, using the default 2HDMC methods. In addition, a number of experimental constraints were tested against, using the most recent version of the relevant public numerical tool. In case an important experimental result had not (yet) been implemented in the tool used, it was explicitly enforced. 

Since the discovery of the $\hobs$, the CMS and ATLAS collaborations at the LHC have frequently updated the measurements of not only its mass but also the signal strengths of its $\gamma\gamma$, $ZZ$, $WW$, $\tau\tau$ and $b\bar{b}$ decay channels \cite{Aad:2015zhl,Khachatryan:2016vau}. The theoretical counterparts of the signal strengths of the assumed $\hobs$ candidate in the 2HDM case being studied ought to be consistent with these measurements. For each analysis, we calculated these observables using the program HiggsSignals \cite{Bechtle:2013xfa}, and required them to lie within $2\sigma$ of the latest (combined) LHC measurements available at the time. Besides the $\hobs$, the masses of the additional Higgs bosons are also strongly constrained by the null results from their direct searches at the Large Electron-Positron (LEP) collider, the TeVatron and the LHC. We used the HiggsBounds code \cite{Bechtle:2013wla} to impose the up-to-date $95\%$ confidence level (CL) limits from these colliders. 

Certain experimental results indirectly constrain the parameter space of the 2HDM also. The first type of such constraints come from the EW precision data, and include the measurement of the $Z$ boson width from LEP \cite{Agashe:2014kda} and the frequently revised ones of the oblique parameters S, T and U \cite{Agashe:2014kda,Tanabashi:2018oca}. We required consistency of the model predictions, calculated by 2HDMC, with the former within 2$\sigma$ of the mean value and with the latter at the 95\% CL. The second type come from $B$-physics. The predictions for a number of these observables for each scanned point were calculated using the SuperIso \cite{Mahmoudi:2008tp} program, and required to agree with the 95\% CL limits suggested in the program's manual (unless specified otherwise).

For the points meeting all the above requirements, we subsequently calculated the tree-level cross sections for the production of various possible di-Higgs states at the 13\,TeV LHC. For the EW production, $q\bar q^{(\prime)}\to h_i h_j$, with $h_{i,j} = (h,\;H,\;A,\;\hpm)$, we used the 2HDMC model \cite{Eriksson:2009ws} with MadGraph5\_aMC@NLO \cite{Alwall:2014hca}. Cross sections for the QCD-induced processes, $b\bar{b}\to h_i h_j$ ($gg\to h_i h_j$), for neutral final states only, were also computed using MadGraph (based codes \cite{Hespel:2014sla}). 

\section{The SM-like $H$ case}

\begin{table}[t]
\begin{center}
\renewcommand{\arraystretch}{1.2}
\begin{tabular}{ c | c c c c c c | c }
BP & $m_h$ & $m_A$ & $m_{H^\pm}$  & $\sin(\beta-\alpha)$ & $m_{12}^2$ & $\tan\beta$ & $\cos\alpha/\sin\beta$ \\ \hline \hline
1 & 54.2 & 33.0 & 95.9 & $-0.11590$ & 118.3 & 9.0947 &  $-6.7\times 10^{-3}$ \\ 
2 & 22.2 & 64.9 & 101.5 & $-0.046960$ & 10.6 & 22.114 & $-1.8\times 10^{-3}$ \\  \hline
3 & 14.3 & 71.6 & 107.2 & $-0.061929$ & 2.9 & 16.307 & $-7.2 \times 10^{-4}$ \\ \hline
4 & 27.5 & 117.8 & 86.8 & $-0.14705$ &  44.5 &  6.8946 & $-3.6\times 10^{-3}$ \\
5 & 63.3 & 129.2 & 148.0 & $-0.048763$ &  173.1 &  20.660 & $-4.2\times 10^{-4}$ \\ 
\end{tabular}
    \caption{Parameter values corresponding to the five benchmark points. All masses are in GeV.}
	\label{tab:BPs}
\end{center}
\end{table}

We first consider the case in which the heavier CP-even scalar $H$ is identified with the SM-like Higgs boson, by fixing its mass to 125\,GeV and requiring its signal strengths to be consistent with those of the $\hobs$. This implies that $h$ is by definition lighter than 125\,GeV. Two studies pertaining to this case, concentrating on two interesting scenarios realisable in specific 2HDM parameter space regions, were carried out, following for each study the methodology explained above. In Table \ref{tab:BPs} we show the parameter combinations for five benchmark points (BPs) selected from the points collected for the two studies, discussed in detail below. Of particular significance here is the parameter $\sin(\beta-\alpha)$, which typically tends towards very small (negative) values. The reason is that the $g_{HAZ}$ and $g_{H\hpm\wmp}$ ($g_{hAZ}$ and $g_{h\hpm\wmp}$), which are of relevance to the two scenarios considered, are both proportional to $\sin(\beta-\alpha)$ ($\cos(\beta-\alpha)$). Large values of $\sin(\beta-\alpha)$ would boost the decay of $H$ into $AZ^{(*)}$, which is tightly constrained by the LHC searches. This parameter (and its interplay with $\tan\beta$) also governs the couplings of $H$ to the fermions as well as to pairs of other Higgs bosons, and the condition on it to be SM-like pushes the model into the `alignment limit', $\sin(\beta-\alpha) \to 0$ \cite{Ferreira:2012my,Bernon:2015wef}. 

\subsection{Pair-production of light $h$ and $A$}

\begin{figure}[t!]
\begin{center}
\begin{tabular}{cc}
\hspace*{-0.3cm}\includegraphics[scale=0.42]{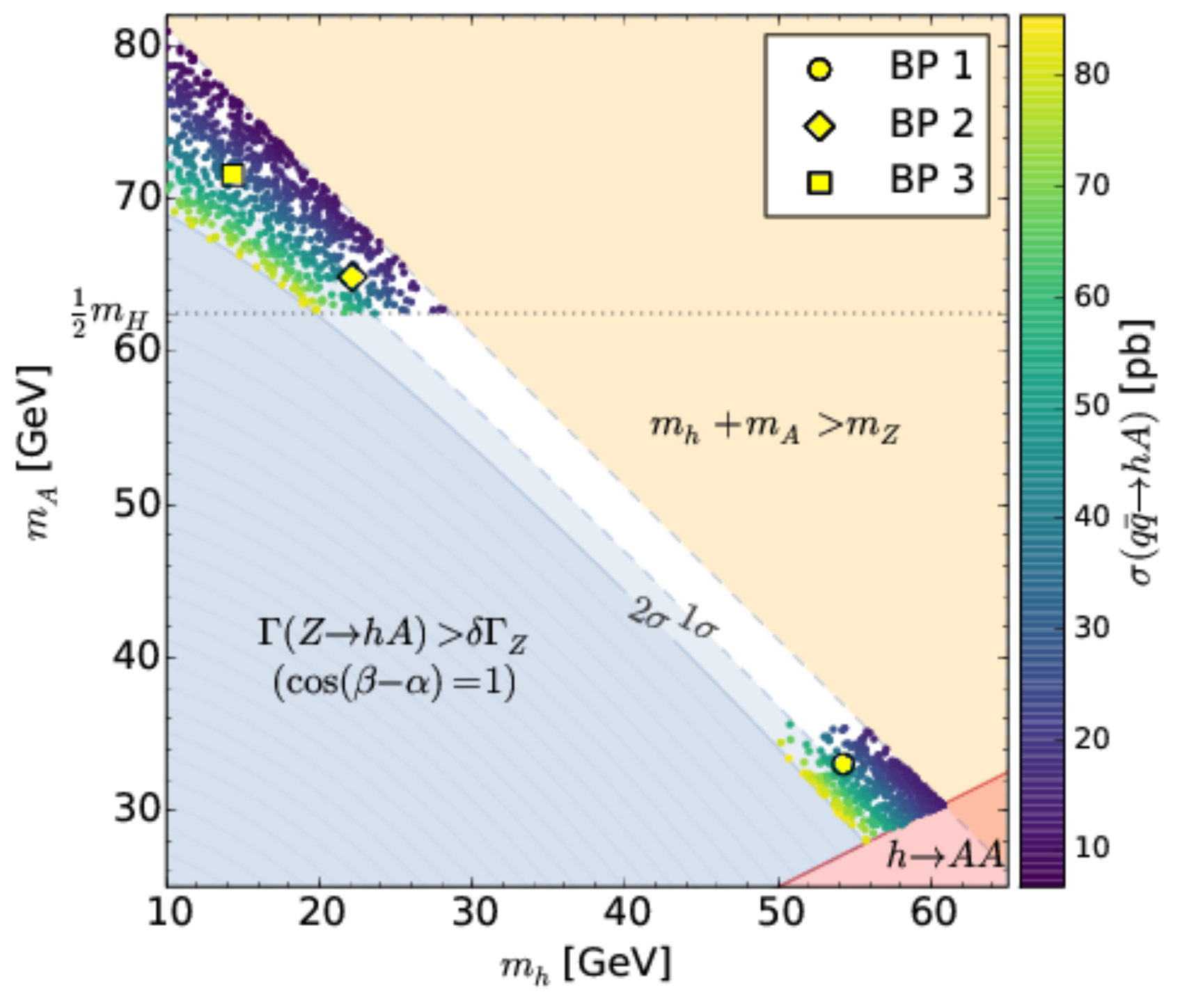} &
    \includegraphics[scale=0.40]{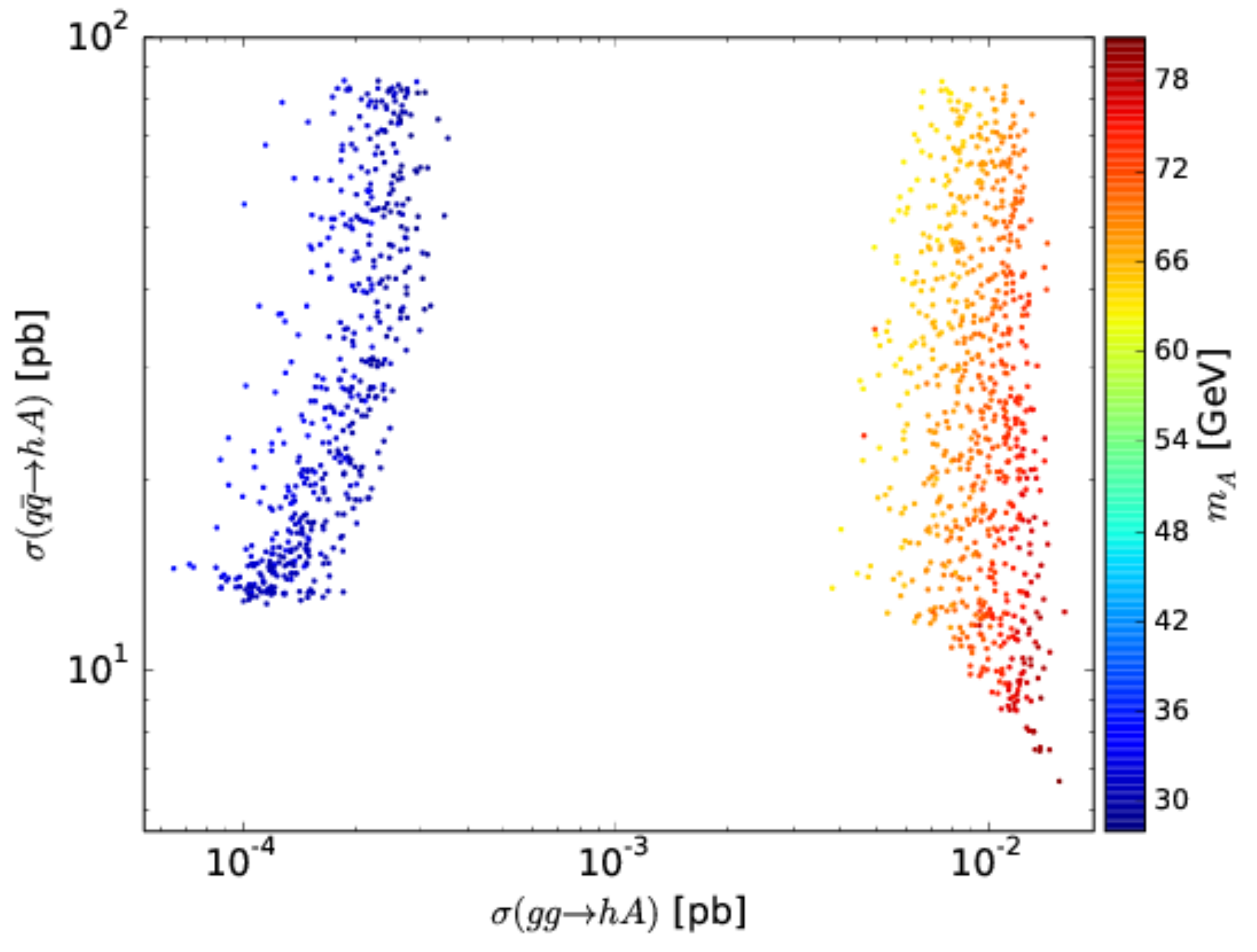}
\end{tabular}
    \caption{Left: Successful points from the scan that additionally lie within $1\sigma$ (lighter) and $2\sigma$ (darker) of  the experimental uncertainty on the $Z \to h A$ partial width. The colour heat map corresponds to the total cross section for the $q\bar{q}\to hA$ process at $\sqrt{s}=13$\ TeV. Right: Comparison of the cross sections for the EW and QCD-induced productions of $hA$ pairs, with the colour heat map corresponding the mass of $A$.}
    \label{fig:qqvsgg}
\end{center}
\end{figure}

In the scenario where, in addition to the $h$, the pseudoscalar $A$ is light enough such that $m_h+m_A<m_Z$, their pair-production via a resonant $Z$ should become kinematically available. However, the Landau-Yang theorem \cite{Landau:1948kw,Yang:1950rg} prohibits the on-shell production of a $Z$ boson in gluon-gluon scattering. The EW process $q\bar{q} \to Z \to hA$, on the other hand, faces no such limitation, and could therefore have a substantial cross section near the $Z$ boson production threshold. In order to investigate this possibility, we scanned the masses of the non-SM Higgs bosons in the ranges \cite{Enberg:2016ygw}
\begin{center}
		$m_h= 10$ -- 80\,GeV\,;~~$m_A= 10$ -- $M_Z - m_h$ \,GeV\,;~~$m_{\hpm} =90$ -- 500\,GeV\,,\\
\end{center}
along with the other three free parameters, the default ranges of which have been noted earlier.

The alignment limit (interpreted as $\cos(\beta-\alpha)\to 1$) maximises the $g_{hAZ}$ coupling and, in turn, the $Z \to hA$ partial width, which is subject to stringent limits from the LEP measurement \cite{Agashe:2014kda}. In the left frame of Fig.\ \ref{fig:qqvsgg} we show the successful points from the scan, that additionally lie within the experimental uncertainty on this partial width, at the $1\sigma$ (lighter) and $2\sigma$ (darker) levels, assuming $\cos(\beta-\alpha)=1$.  There are two distinct regions with a large density of points in the figure. One of them lies near the top left corner, corresponding to $m_A > m_h$, and cuts off sharply at $m_A=m_H/2$, when the experimentally disfavoured $H\to A A$ decay becomes available. The probability of this decay can be reduced by sufficiently suppressing the $g_{HAA}$ coupling, which is what causes the points with $m_h > m_A$ to  reappear near the lower right corner of the figure. This region also gets truncated when the $h\to A A$ decay, which is excluded by experiment \cite{Schael:2006cr}, opens up owing to $m_h > 2m_A$. BPs 1, 2 and 3, defined in Table\,\ref{tab:BPs}, have been highlighted in yellow in the figure, and the total cross section for the $q\bar{q}\to hA$ process is depicted by the colour heat map. 

\begin{table}
\begin{center}
\begin{tabular}{c | c c | c c c c | c c c}
\multirow{2}{*}{BP} & \multicolumn{2}{c|}{$\sigma$ [fb]} & \multicolumn{4}{c}{BR$(h\to ...)$ [\%]} & \multicolumn{3}{|c}{BR$(A\to ...)$ [\%]} \\
& $q\bar{q}\to hA$ & $gg\to hA$ & $Z^* A$ & $b\bar{b}$ & $\gamma\gamma$ & $\tau\tau$ & $Z^* h$ & $b\bar{b}$ & $\tau\tau$ \\ \hline
1 & 41.2 & $1.5 \times 10^{-4}$ &94 & 5 & $<1$ & $<1$ & 0 & 86 & 7\\ 
2 & 34.4 & $7.2 \times 10^{-3}$ & 0 & 83 & 3 & 7 & 86 & 12 & 1\\ 
\end{tabular}
\caption{Signal cross sections for the EW and QCD-induced $hA$ production. Also given are the dominant BRs of $h$ and $A$ for the BPs corresponding to this scenario.}
\label{tab:hA}
\end{center}
\end{table}

From the right frame of Fig.\,\ref{fig:qqvsgg} it is evident that the cross section for the EW production of $hA$ can exceed that for the QCD production by a few orders of magnitude. Table \ref{tab:hA} shows that for BP1, with $m_h>m_A$, the difference between the two cross
sections is much more pronounced compared to that for the BP2, with $m_A>m_h$. Also included in the table are the branching ratios
(BRs) of $h$ and $A$ in their most dominant decay modes. $AZ^*$ is the primary decay channel of $h$ for BP1, and for BP2, $A$ decays predominantly to $hZ^*$. As $b\bar{b}$ is the preferred decay mode of both the resulting $A$ and $h$, for BP 1 and 2, respectively, final states like $Z^*b\bar{b}b\bar{b}$ and $Z^*b\bar{b}\tau\tau$ could be crucial signatures of this scenario at the LHC.    

\subsection{$H^\pm$ production along with a fermiophobic $h$}

As pointed out above, the coupling $g_{h\hpm\wpm}\sim\cos(\beta-\alpha)$ also gets maximised in the alignment limit, resulting in an enhancement in both the $\sigma(pp\to W^{\pm*}\to h H^\pm)$ \cite{Haisch:2017gql} and the BR$(H^\pm\to h W^\pm) $ \cite{Arhrib:2016wpw}. In fact, below the $t\bar{b}$ threshold, the BR($\hpm \to h W^\pm $) can approach unity in this limit. Therefore, to explore this second plausible scenario in this 2HDM case, the Higgs boson masses were scanned in the ranges
\begin{center}
		$m_h=$ 10 -- 120\,GeV\,;~~$m_A=$ 10 -- 500\,GeV\,;~~$m_{\hpm}=$ 80 -- 170\,GeV\,,\\
\end{center}
thus allowing both $h$ and $A$ to take up larger values than in the first scenario. 

\begin{figure}[t!]
\begin{center}
\includegraphics[width=0.48\textwidth]{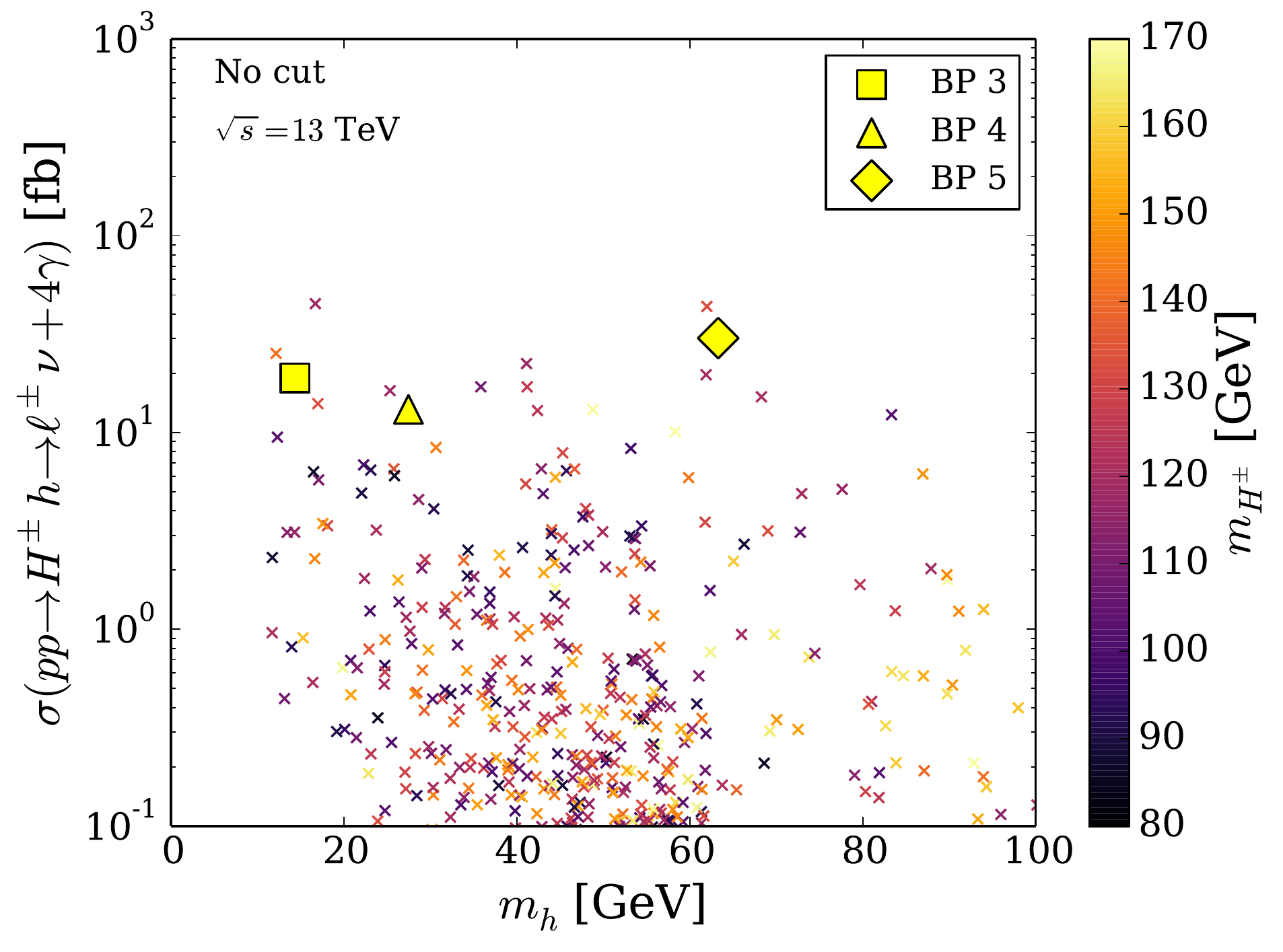}
\includegraphics[width=0.48\textwidth]{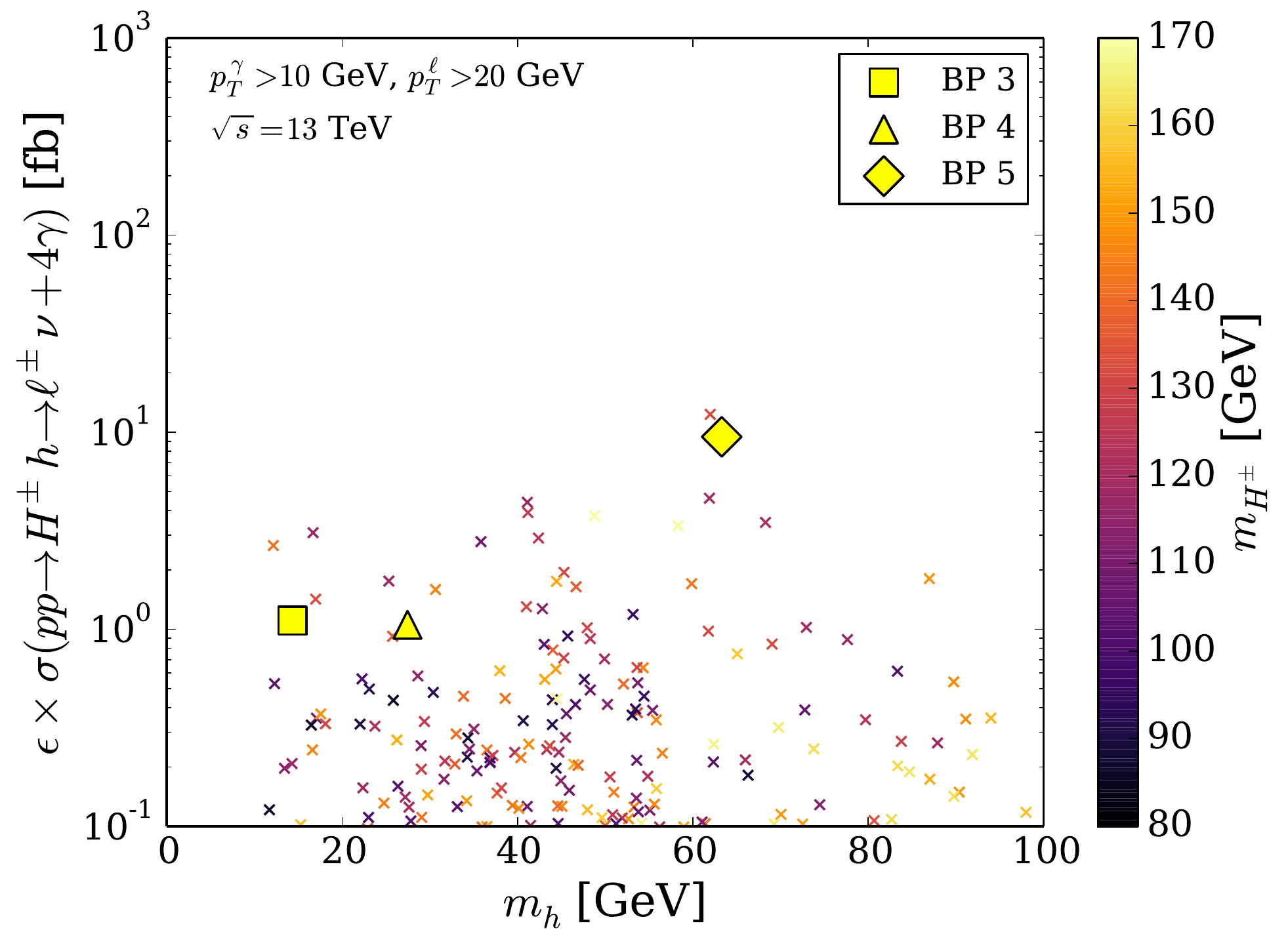}
\caption{Cross section for the process $pp \to \ell^\pm\nu+4\gamma$ before imposing the selection cuts (left) and after multiplying with the selection efficiencies (right), with the colour heat map depicting the mass of $\hpm$. BPs 3, 4 and 5 are highlighted in yellow.}
\end{center}
\label{fig:sigma-wh}
\end{figure}

In the Type-I 2HDM, the couplings of $h$ to all the fermions are proportional to $\cos(\alpha)/\sin\beta$. Since $\cos\alpha = \sin\beta\sin(\beta-\alpha)+\cos\beta\cos(\beta-\alpha)$, these couplings can vanish for certain combinations of $\sin(\beta-\alpha)$ and $\tan\beta$, making the $h$ highly fermiophobic \cite{Akeroyd:2007yh}. However, the effective coupling of $h$ to two photons, which is dominated by {\it t}-quark, $\hpm$ and $\wpm$ loop contributions, may not be suppressed for the same combinations, leading to a sufficient partial width and a sizeable BR for the $h\to \gamma\gamma$ decay, especially when the $h\to WW^*$ decay channel is kinematically closed. Consequently, the cross section for the process $pp\to H^\pm h\to W^\pm hh\to \ell^\pm\nu +4\gamma$ (which we calculated as $\sigma(qq'\to H^\pm h)\times {\rm BR}(H^\pm \to W^\pm h)\times {\rm BR}(h\to \gamma\gamma)^2\times {\rm BR}(W^\pm\to\ell^\pm\nu)$) can reach a few tens of femtobarns, especially for $m_h \lesssim 60$\,GeV and $m_\hpm \lesssim 120$\,GeV, as noted in the left frame of Fig.\,\ref{fig:sigma-wh}. 

However, because of the small masses of the decaying Higgs bosons, some of the final state objects are likely to be soft, which could make this a rather challenging channel, even though it benefits from a tiny SM background. We therefore also estimated the selection efficiencies using the most optimal set of cuts for the mass ranges of the Higgs bosons involved \cite{Arhrib:2017wmo}. These cuts cuts include: $p_T^\gamma>10$\,GeV and $p_T^l>20$\, GeV for the transverse momenta of the photon (with the lowest momentum of the four) and the lepton, respectively, $|\eta|<2.5$ for the pseudorapidies of the lepton and all the photons, and isolation $\Delta R = \sqrt{(\Delta\eta)^2+(\Delta\phi)^2} > 0.4$ for all the objects. The signal efficiencies were then determined as $\epsilon=\sigma(\text{cuts})/\sigma(\text{no cuts})$. The right frame of Fig.\,\ref{fig:sigma-wh} illustrates that the cross sections for the scanned points after multiplication with these efficiences, which are functions of $m_h$ and $m_\hpm$, can still reach up to 10\,fb. In Table~\ref{tab:w4gamma} we quote the two BRs that are of relevance for this process and its cross section (which does not take into account the BR$(W^\pm \to \ell^\pm \nu)$ and $\epsilon$), for each of the BPs 4 and 5. (BP3 will be addressed later.) 

We point out here that while $m_A$ was allowed to take up much higher values in our scan for this scenario, it always stays close to $m_\hpm$ in order to satisfy the constraints on the oblique parameters. This has the important phenomenological implication that the BR($A\to hZ^*)$ too can be extremely large in this parameter space region, as also seen in table~\ref{tab:w4gamma}. It can thus be responsible for a substantial cross section for the process $pp\to \hpm A\to \wpm h+Z^*h\to \wpm Z^*+4\gamma$ as well, that may potentially lead to another four-photon signature of this model.  

\begin{table}[t!]
\begin{center}
\begin{tabular}{c | c c c | c c }
\multirow{2}{*}{BP} & \multicolumn{3}{c|}{BR [\%]}& \multicolumn{2}{c}{$\sigma$ [fb]}  \\
& $H^\pm \to W^\pm h$ & $A \to Z^* h$ & $h\to\gamma\gamma$ & $W^\pm+4\gamma$ & $W^\pm Z^*+4\gamma$  \\\hline
4 & 98 & 94 & 16 & 61.5 & 7.4 \\
5 & 100 & 98 & 71 & 141.4 & 55.7 \\
\end{tabular}
\caption{Cross sections for the considered EW processes, and the dominant BRs of $H^\pm$, $h$, and $A$, for BPs 4 and 5.}
\label{tab:w4gamma}
\end{center}
\end{table}

\subsection{A threefold 4-photon signature}

A large cross section for the $pp\to hH^\pm $ process in the second scenario above generally prefers slightly larger $m_A$ compared to the first one (so that it is not completely dominated by the $pp\to A\hpm$ process). However, there is an intermediate range, which the BP3 lies in, for which the $hA$ production from a resonant $Z$ boson is also allowed. Owing to the highly fermiophobic nature of $h$ and a significant BR($h\to \gamma\gamma)$, this $hA$ pair could be probed in the $Z^*\gamma\gamma\gamma\gamma$ state with a vanishing background. According to Table~\ref{tab:BP3}, the (pre-selection) cross section for this signal lies at the femtobarn level, as do those of $W^\pm+4\gamma$ and $W^\pm Z^*+4\gamma$. Concerning the latter two signatures, a large $m_{\hpm}-m_h$ difference for this BP should lead to a harder spectrum for a lepton from $W^\pm$ decay, so that the selection efficiency might be somewhat higher than for the BPs 4 and 5. 

 \begin{table}[t!]
\begin{center}
\begin{tabular}{| c | c |}
\hline
\multicolumn{2}{|c|}{BP 3} \\ \hline \hline
$\sigma (q\bar{q}\to hA \to Z^*+4\gamma)$ & 1.64 fb \\ 
$[\sigma (gg\to hA \to Z^*+4\gamma)]$ & $[5.7 \times 10^{-4}$ ${\rm fb}]$ \\ \hline 
$\sigma(q\bar{q}\to H^\pm h \to W^\pm+4\gamma)$ & 88.8 fb \\ 
$\sigma(q\bar{q}\to H^\pm A \to W^\pm Z^*+4\gamma)$ & 26.8 fb \\ \hline
BR$(H^\pm \to W^\pm h)$ & 100 \%  \\ 
BR$(A\to Z^*h)$ & 90 \% \\ \hline
BR$(h\to \gamma\gamma)$ & 24 \% \\
BR$(h\to b\bar{b})$ & 60 \% \\ \hline
\end{tabular}
\caption{Cross sections for the three EW processes that can potentially yield 4-photon signatures, and the dominant BRs of $H^\pm$, $A$ and $h$, for BP3.}
\label{tab:BP3}
\end{center}
\end{table}

\section{The SM-like $h$ case}

When the lighter $h$ is identified with the $\hobs$, $H$ is essentially allowed to be very heavy. In our numerical scan, though, we restricted the additional Higgs boson masses to the ranges \cite{Enberg:2018pye}
\begin{center}
		$m_H=$ 150 -- 750\,GeV\,;~~$m_A=$ 50 -- 750\,GeV\,;~~$m_{\hpm}=$ 50 -- 750\,GeV\,,\\
\end{center}
in order that sufficient phase space is available for the production of various di-Higgs states. A crucial feature of this case is the possibility of the decays of the heavier Higgs bosons into (pairs of) the 125\,GeV $h$. This means that their probes in the well-established experimental analyses for the $\hobs$ can strongly compliment their direct searches in the SM decay channels. Therefore, in contrast with the first case, here the $g_{hAZ}$ and $g_{h\hpm\wmp}$ couplings (besides, naturally, the triple-Higgs couplings $g_{hhH}$, $g_{hHH}$, $g_{hAA}$ and $g_{h\hpm\hmp}$) ought to be sufficiently small for the relevant decays to evade the latest collider constraints. Since both these couplings are proportional to $\cos(\beta-\alpha)$, this parameter is confined to values close to zero, which corresponds to the `decoupling limit' \cite{Gunion:2002zf,Branco:2011iw} of the model. Also in the scan for this case, we imposed the following updated 95\% CL limits on the $B$-physics observables: 
\begin{itemize}
\item ${\rm BR}(B\to X_s \gamma) $: $3.32\pm0.15 \times 10^{-4}$~\cite{Amhis:2016xyh},
\item ${\rm BR}(B_u\to \tau^\pm \nu_\tau) $: $1.06\pm0.19 \times 10^{-4}$~\cite{Amhis:2016xyh},
\item ${\rm BR}(B_s \to \mu^+ \mu^-)$: $3.0\pm 0.85\times 10^{-9} $~\cite{Aaij:2017vad}.
\end{itemize}

\begin{figure}[tb!]
\includegraphics[angle=0,width=0.48\textwidth]{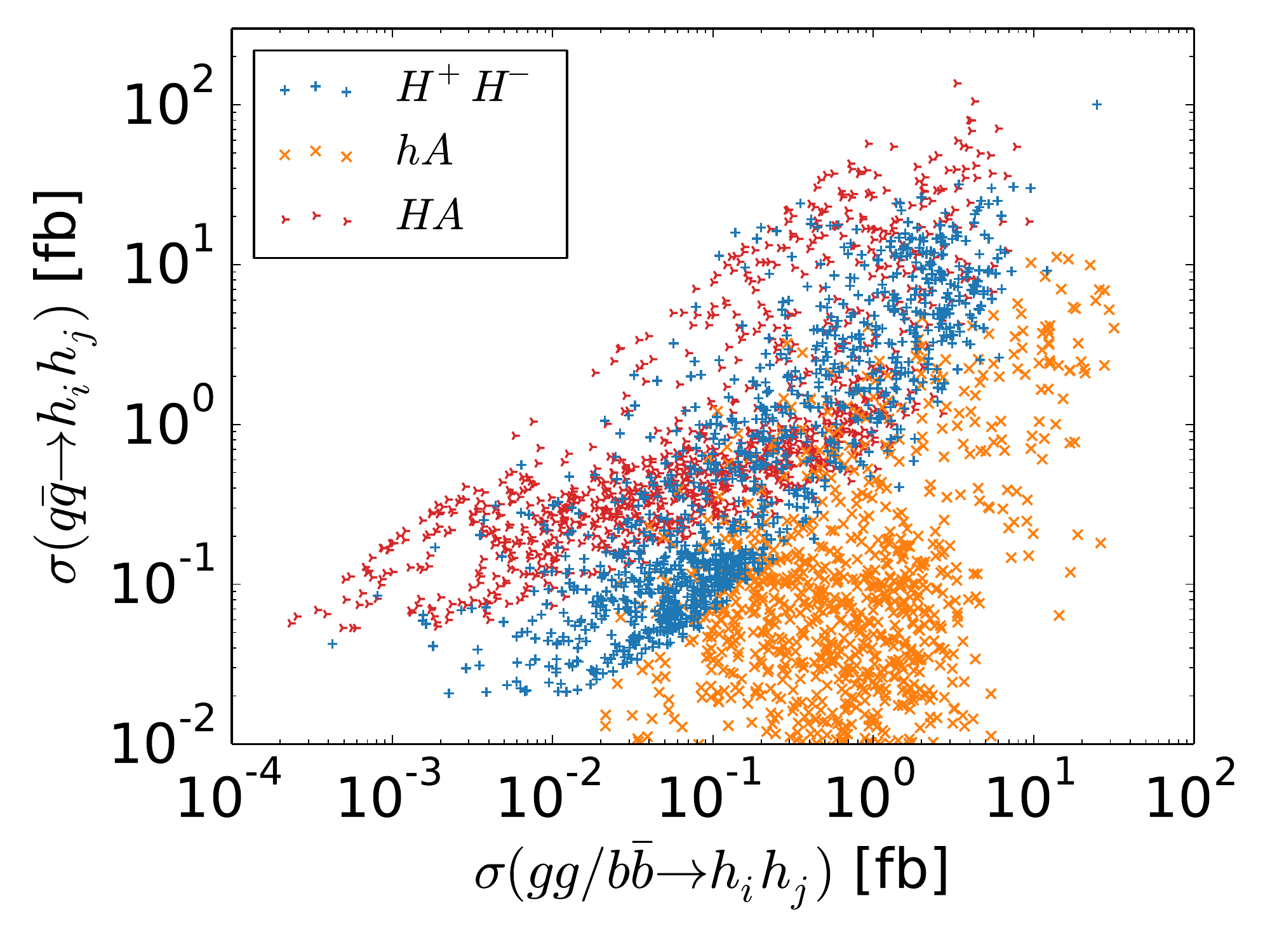}
\hspace*{0.5cm}\includegraphics[angle=0,width=0.48\textwidth]{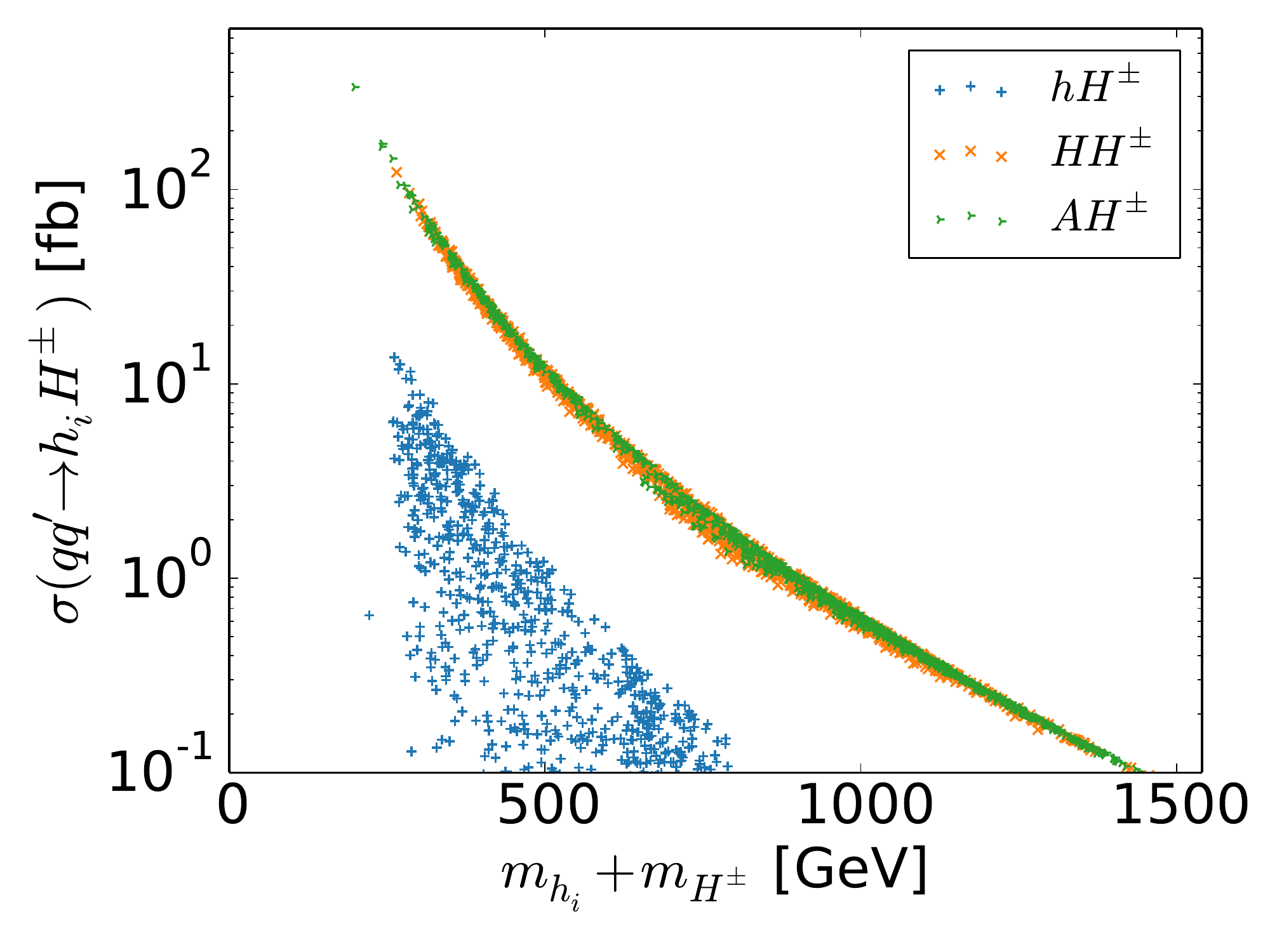}
\caption{Left: Neutral di-Higgs states for which the cross sections for $qq^{\prime}$ production can exceed those for $gg/b\bar{b}$-initiated processes. Right: Cross sections for the three possible charged di-Higgs states.}
\label{fig:2bfs}
\end{figure}

\subsection{Neutral di-Higgs states}

For the points obtained from the scan, we calculated the cross sections for the EW production of all possible neutral di-Higgs states, i.e., for all $q\bar q^{(\prime)}\to h_i h_j$ processes with $h_{i,j} = (h,\;H,\;A)$. As noted earlier, such states may also be produced in gluon-initiated processes involving heavy quark loops, as well as in $b\bar{b}$-fusion, the combined cross sections for which were calculated as well, for comparison. Both these types of cross sections are shown in the left frame of Fig.~\ref{fig:2bfs}. We see that for the $h A$, $H A$ and $H^+H^-$ states the EW production can dominate the $gg/b\bar{b}$ production, with the cross sections for the first reaching above 10\,fb for $h A$, and up to about 100\,fb for $H A$ and $H^+H^-$. For the remaining neutral 2BFSs, however, the EW cross sections did not exceed 1\,fb for any of the points, and they are thus not shown here.

\subsection{Charged di-Higgs states}

Cross sections for the three possible pairings of $H^\pm$ with one neutral Higgs state, which can only be produced electroweakly, are shown in the right frame of Fig.~\ref{fig:2bfs}, as functions of the sums of their masses. The cross section for the $h\hpm$ pair can only reach about 10\,fb for a handful of points, but those for the $HH^\pm$ and $AH^\pm$ states can exceed 100\,fb for smaller masses. The main contribution to these cross sections comes from the $s$-channel processes mediated by a $W^{\pm *}$, so that the couplings of relevance are $g_{H\hpm\wmp}$, which is proportional to $\sin(\beta-\alpha)$, and $g_{A\hpm\wmp}$, respectively. The fact that almost all the points collected in the scan belong in the decoupling limit, with $\cos(\beta-\alpha) \to 0$, is therefore responsible for the large $H\hpm$ cross sections. The $AH^\pm W^\mp$ coupling, on the other hand, does not depend on $\sin(\beta-\alpha)$, so the substantial cross sections achievable for the production of $A\hpm$ (which were also seen in the SM-like $H$ case) are determined almost entirely by the final state kinematics.

\subsection{3-body final states and the triple-Higgs couplings}

\begin{figure}[t!]
\begin{center}
\includegraphics[angle=0,width=0.48\textwidth]{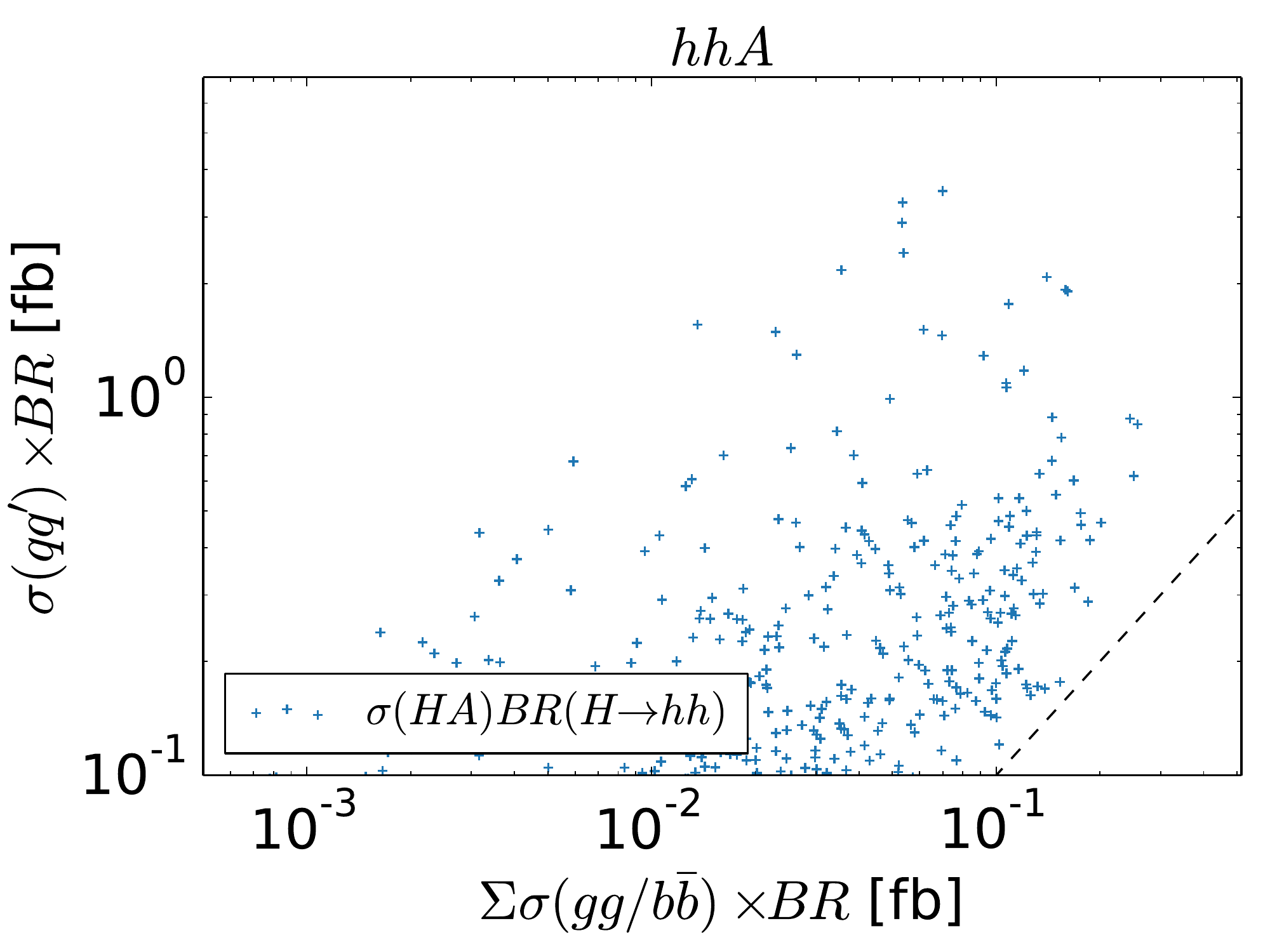}
\hspace*{0.5cm}\includegraphics[angle=0,width=0.48\textwidth]{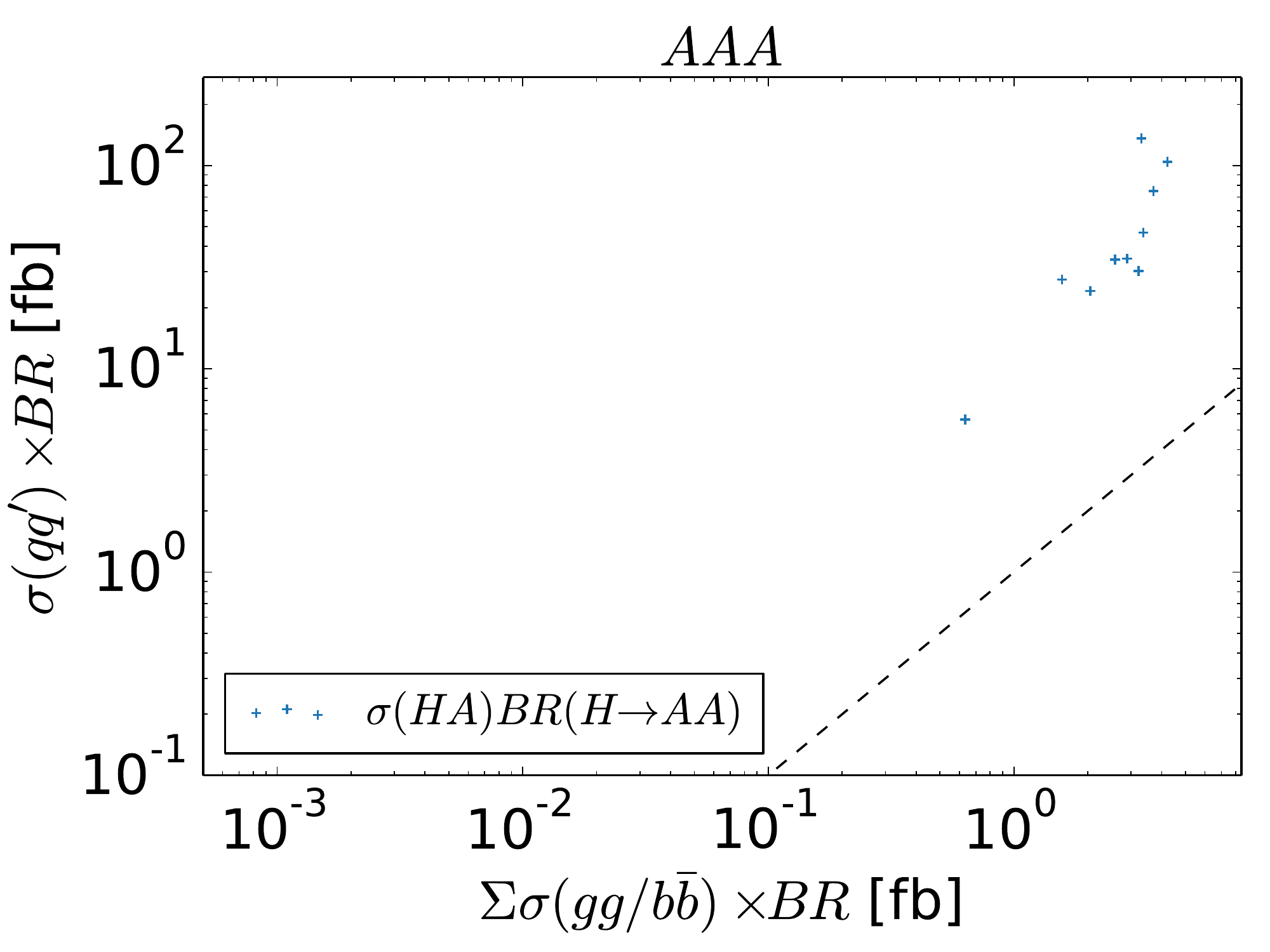}\\
\includegraphics[angle=0,width=0.48\textwidth]{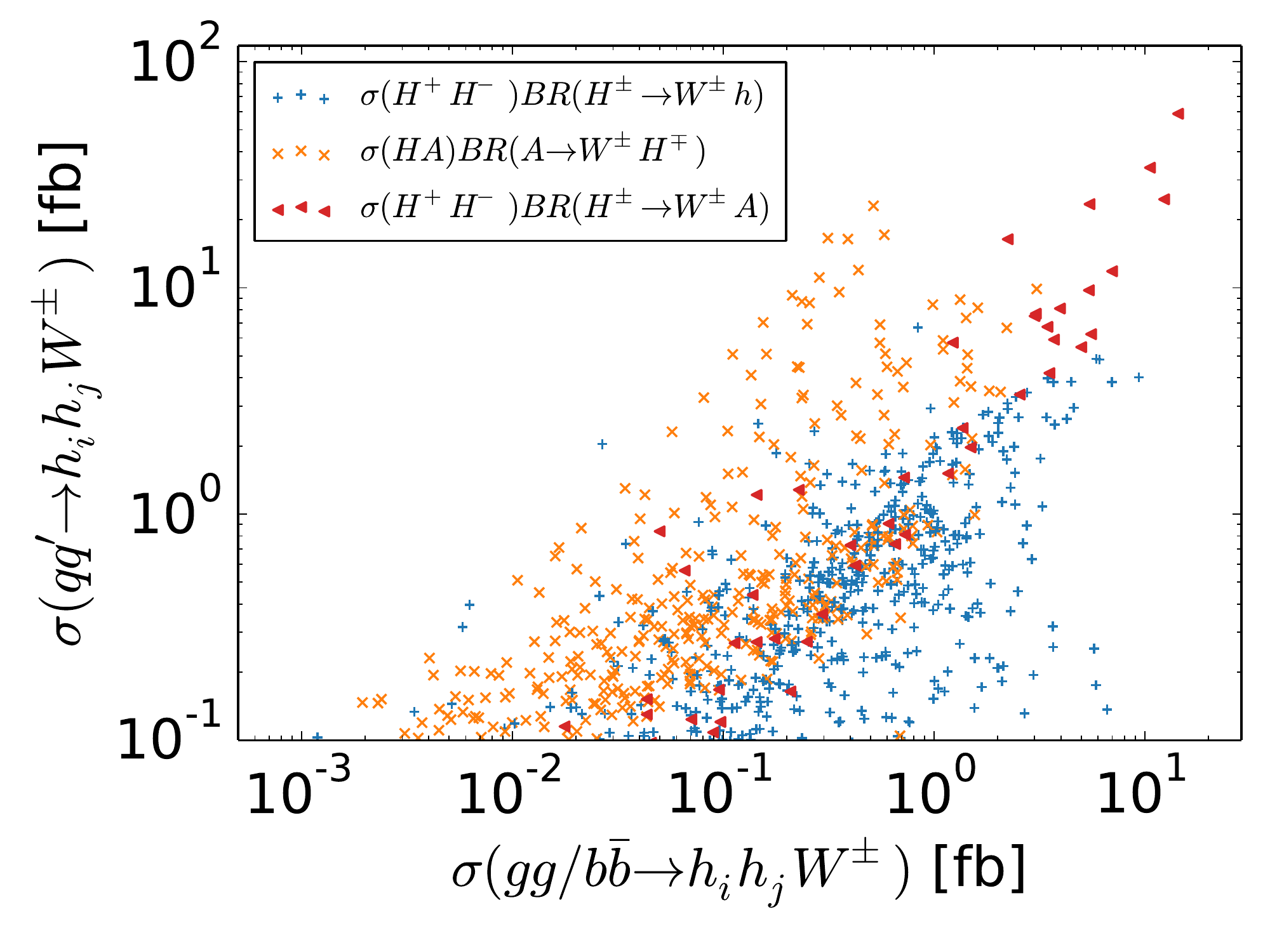}
\hspace*{0.5cm}\includegraphics[angle=0,width=0.48\textwidth]{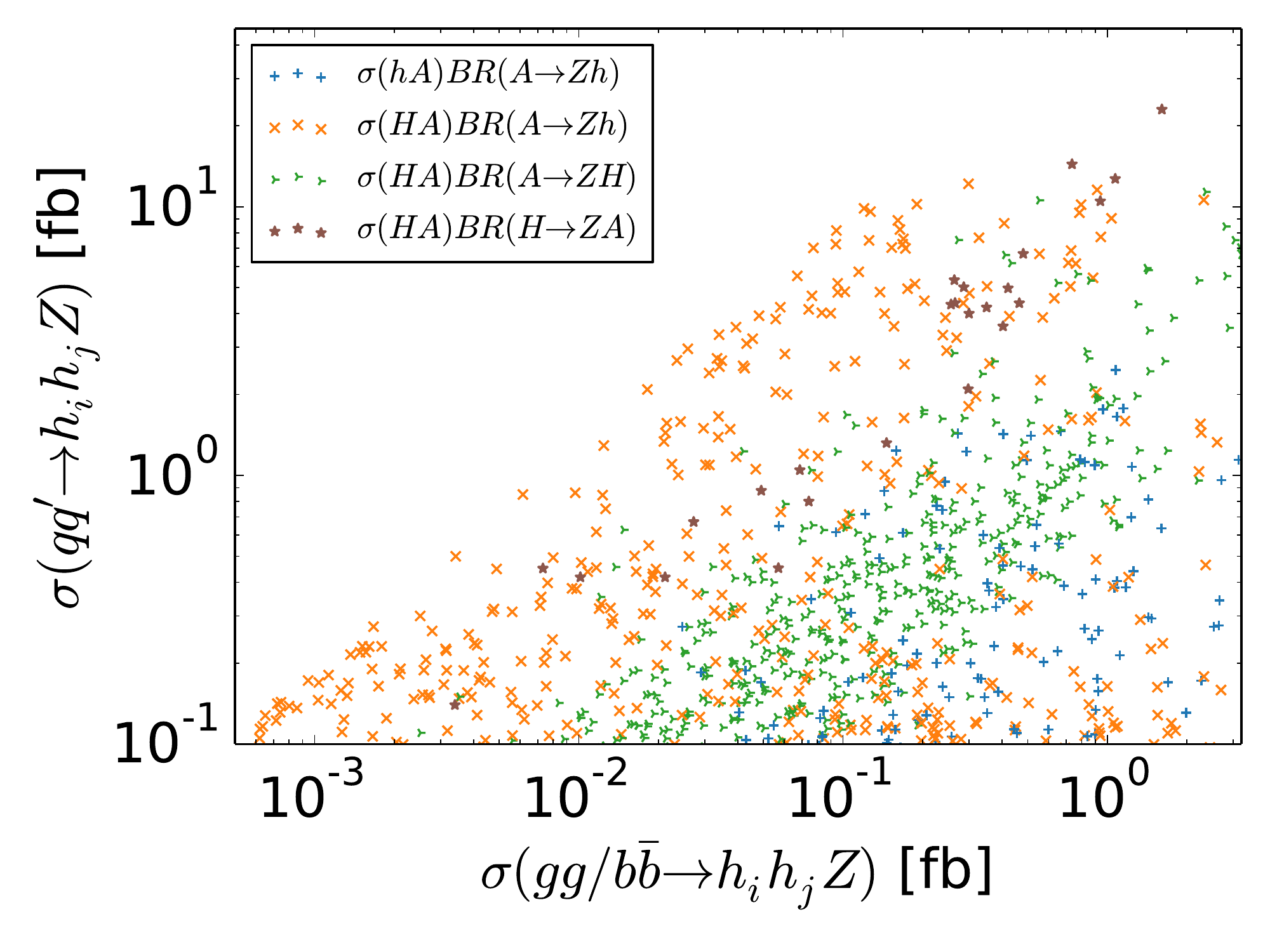}
\caption{Neutral 3-body states for which $gg/b\bar{b}$ production dominates over $qq'$-initiated production. Top frames show the Higgs-only states, while the bottom frames shows those containing two Higgs bosons and one $\wpm$ (left) or one $Z$ (right) boson.}
\label{fig:neutral3bfs}
\end{center}
\end{figure}

We further explored the phenomenological prospects of final states containing either three Higgs bosons or a Higgs boson pair accompanied by a gauge boson, which can result from the decays of one of the two partners in a di-Higgs states. We estimated the yields for all such 3-body states, $h_i + h_j + h_k/V_k$, with $h_i = (h,\;H,\;A,\;H^\pm)$ and $V = (W^\pm,\;Z)$, by multiplying the cross section for a given di-Higgs state with the appropriate BR. All possible on-shell decays of the heavier Higgs states were taken into account, with the exception of they $H\to H^+ H^-$ decay, which is not available for any of the collected points, as the condition $m_H>2m_{H^\pm}$ is never met. 

The resulting neutral tri-Higgs states for which the cross sections for the EW production exceed those for the QCD production are shown in the top frames of Fig.~\ref{fig:neutral3bfs}. These include $hhA$ and $AAA$, with the latter available only for a few points due to the fact that the $H\to AA$ decay is open for a small portion of the scanned parameter space. Note that there might be several possible processes leading to the same 3-body state, but only those giving the maximal cross section for that state are presented in this and the subsequent frames. In the bottom left frame are similarly shown the 3-body states containing one $W^\pm$ and two Higgs bosons (one of which is evidently always the $\hpm$). These states, with EW production dominant, include $h\hpm\wmp$ and $A\hpm\wmp$, with two processes contributing to the second state. The cross section for this state can be a few tens of femtobarns when it emerges from the $\hpm\to A\wpm$ decay in the $H^+H^-$ di-Higgs state, due to the fact that this is the strongest decay mode of $\hpm$, when kinematically allowed. The bottom right frame similarly illustrates the EW cross sections, all of which can reach up to 10\,fb, for 3-body states consisting of one $Z$ and two Higgs bosons, namely $hhZ$, $hHZ$, $HHZ$ and $AAZ$.

\begin{figure}[t!]
\begin{center}
\includegraphics[angle=0,width=0.48\textwidth]{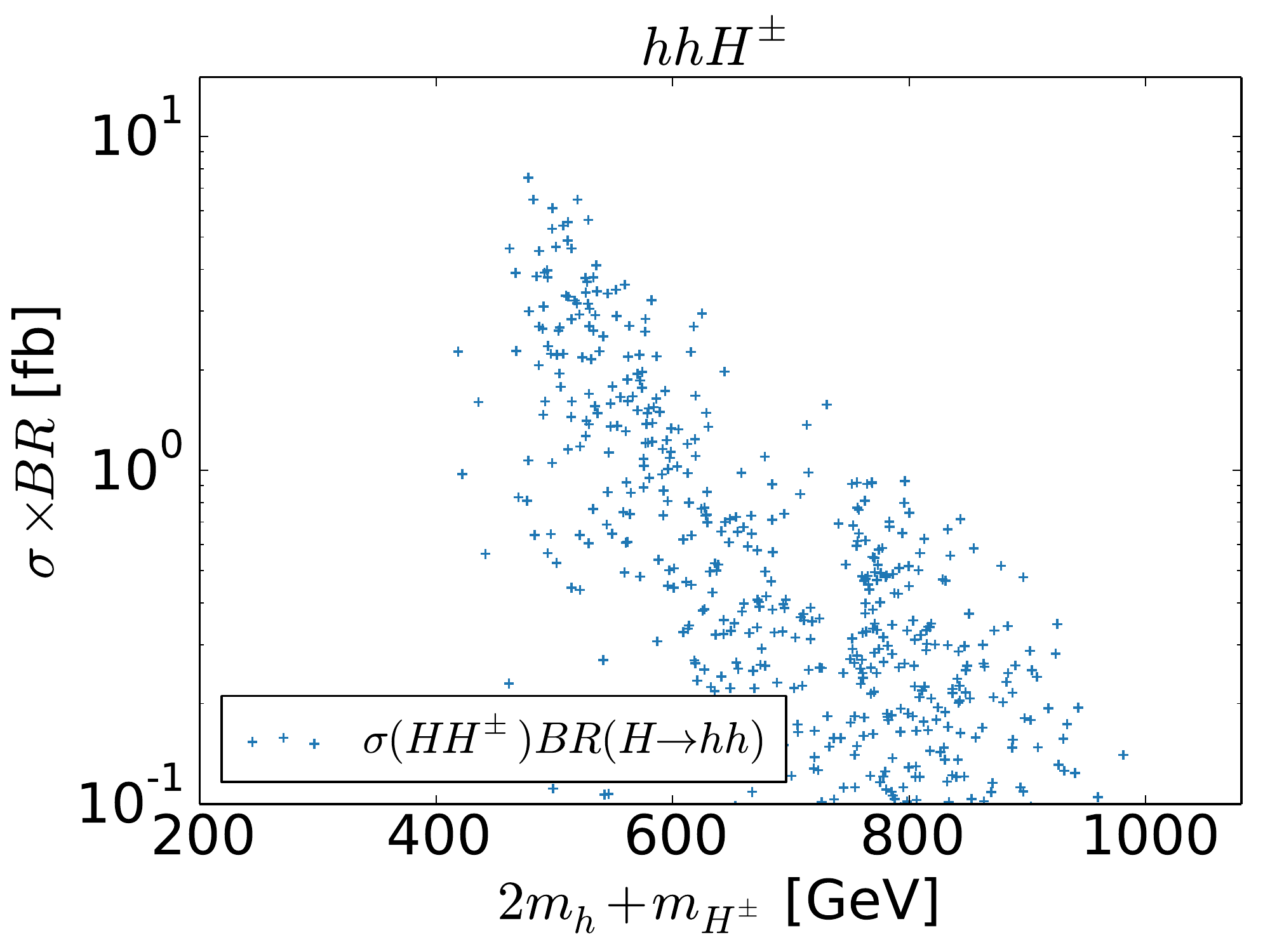}
\hspace*{0.5cm}\includegraphics[angle=0,width=0.48\textwidth]{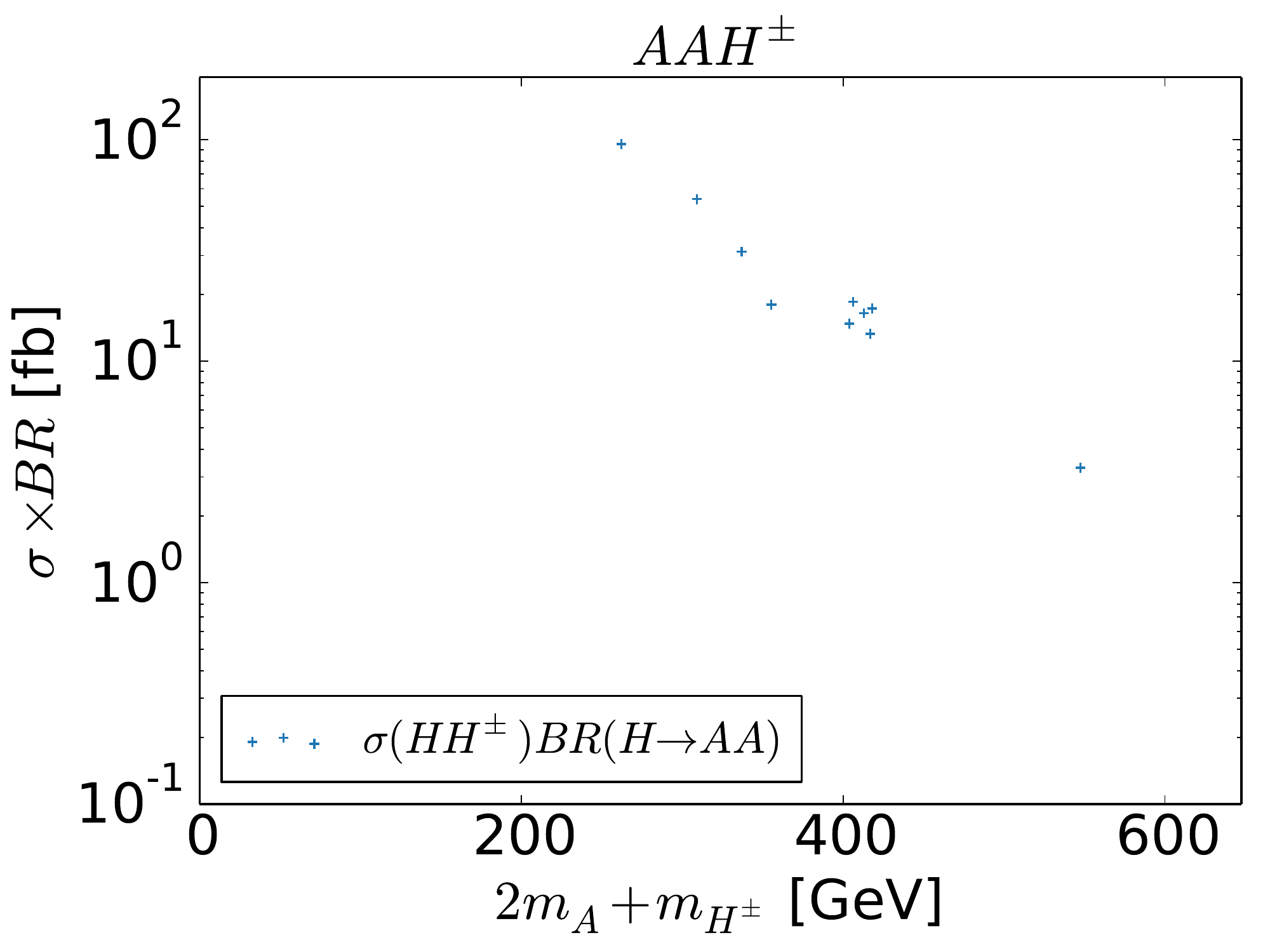}\\
\includegraphics[angle=0,width=0.48\textwidth]{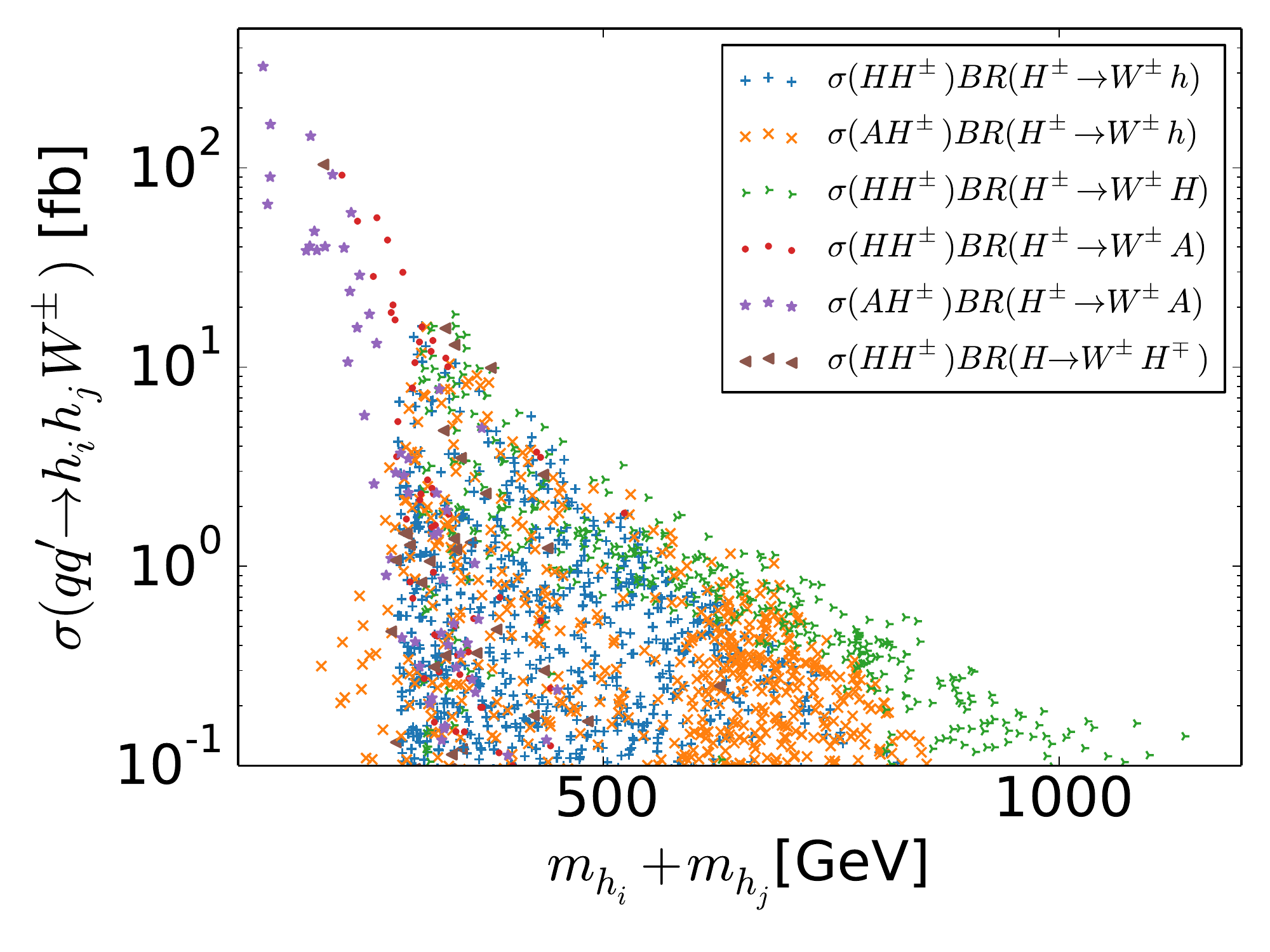}
\hspace*{0.5cm}\includegraphics[angle=0,width=0.48\textwidth]{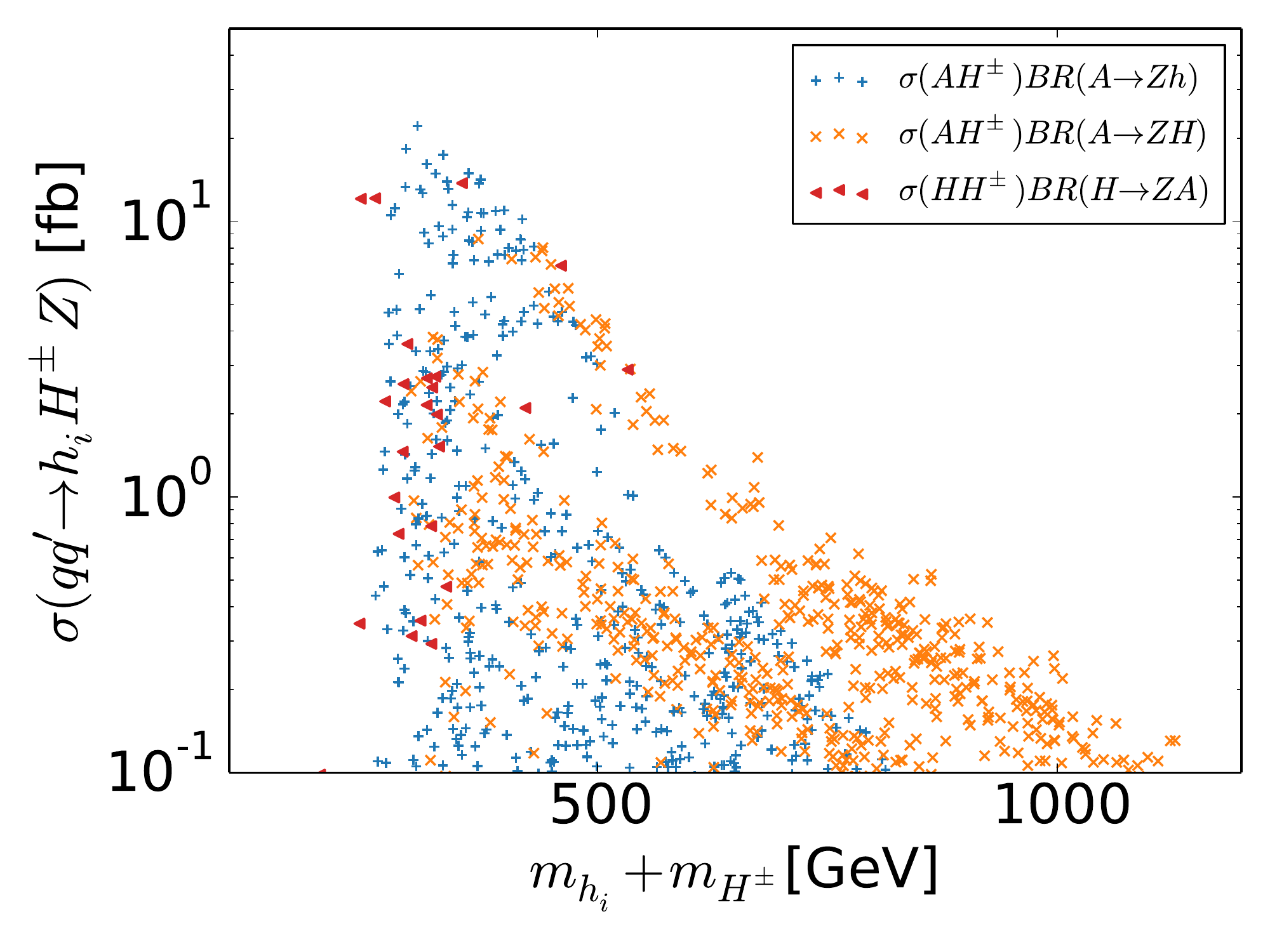}
\caption{Charged 3-body states for which $qq'$-initiated production yields a cross section of at least 0.1\,fb. Top frames show the Higgs-only states, while the bottom frames shows those containing two Higgs bosons and one $\wpm$ (left) or one $Z$ (right) boson.}
\label{fig:charged3bfs}
\end{center}
\end{figure}

Allowing one of the Higgs bosons in a charged di-Higgs state to decay leads to charged 3-body states. From among all possible such states, only the ones for which the production cross section exceeds 0.1\,fb for at least one point from the scan are shown in Fig.~\ref{fig:charged3bfs}. Once again,  all of the possible $h_i\to h_j+h_k/V_k$ decays are represented. In the top frames are the two Higgs-only 3-body states, which include $hh\hpm$ (left) and $AA\hpm$ (right), and in the bottom frames are the Higgs-Higgs-$\wpm$ (left) and the Higgs-Higgs-$Z$ states (right). The reason for very few points with large cross sections involving $H^\pm\to AW^\pm$, $H\to A Z$ and, in particular, $H\to A A$ decays appearing in these frames is that the Higgs boson masses for most of the points do not satisfy the kinematic requirements form. Nevertheless, for the few points that do allow the
$H^\pm\to AW^\pm$ decay, starting with the $A\hpm$ state, the cross section of the resulting 3-body state, $AA\wpm$ (in violet), can exceed 100\,fb.

\begin{figure}[t!]
 \centering\includegraphics[scale=0.9]{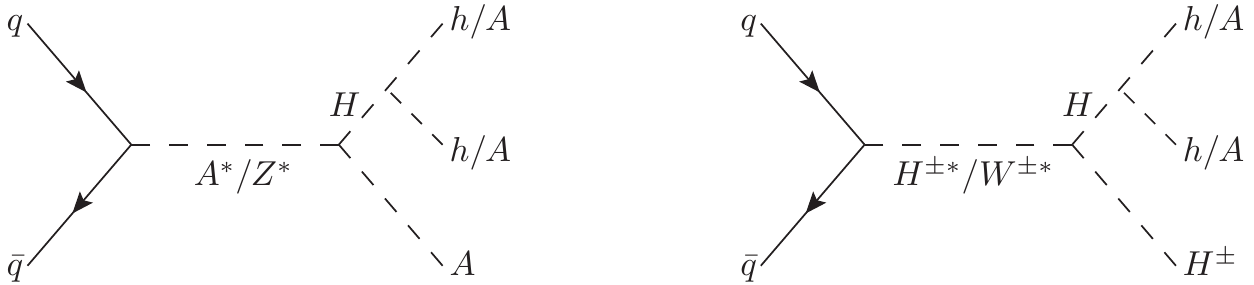}
 \caption{\label{fig:graphs} $s$-channel diagrams contributing to the most dominant Higgs-only three-body states.}
\end{figure}

In the top frames of Figs.~\ref{fig:neutral3bfs} one notices that both the neutral 3-body states with the largest cross sections emerge from the same di-Higgs state, $HA$, which indeed itself has the maximal cross section among all. Likewise, according to Fig.~\ref{fig:charged3bfs} the same $H\hpm$ pair splits into the two charged 3-body states with the highest yield. The dominant diagrams that can lead to these 3-body states are shown in Fig.~\ref{fig:graphs}. Crucially, the same two couplings, $g_{hhH}$ and $g_{HAA}$, are responsible for the decays of $H$ in both these neutral and charged di-Higgs states. The fairly large cross sections (in fact, almost as large as the parent di-Higgs states when $H$ is allowed to decay into $AA$) for the resulting 3-body states could render them accessible at the upcoming LHC Runs. These states may therefore serve as crucial complimentary probes of the triple-Higgs couplings $g_{hhH}$ and $g_{HAA}$, in addition to the established di-Higgs probes where they could potentially enter. 

\acknowledgments{SMo is supported in part through the NExT Institute and the STFC Consolidated Grant ST/L000296/1.
RE, WK and SMo are partially supported by the H2020-MSCA-RISE-2014 grant no. 645722.}


\providecommand{\href}[2]{#2}\begingroup\raggedright\endgroup

\end{document}